\documentclass[12pt]{article}
\usepackage{a4,amsmath,epsfig,amssymb}
\usepackage[latin1]{inputenc}
\usepackage{graphicx}
\usepackage{caption}
\usepackage{subcaption}
\usepackage{feynmp}
\begin{document}
\bigskip\bigskip

\begin{center}
{\large \bf Detailed Study of the $\Lambda_b \to \Lambda_c 
\tau{\bar \nu}_{\tau}$ Decay} 
\end{center}
\vspace{8pt}
\begin{center}
\begin{large}

E.~Di Salvo$^{a,b,}$\footnote{disalvo@ge.infn.it},
F.~Fontanelli$^{b,c,}$\footnote{fontanelli@ge.infn.it} and
Z.~J.~Ajaltouni$^{a,}$\footnote{ziad@clermont.in2p3.fr}
\end{large} 

\bigskip
$^a$ 
Laboratoire de Physique de Clermont - UCA \\
4 Av. Blaise Pascal, TSA60026, F-63178 Aubi\`ere Cedex, France\\

\noindent  
$^b$ 
Dipartimento di Fisica 
 Universit\'a di Genova \\
Via Dodecaneso, 33, 16146 Genova, Italy
 
\noindent  
$^c$
I.N.F.N. - Sez. Genova,\\
Via Dodecaneso, 33, 16146 Genova, Italy \\  

\noindent  

\vskip 1 true cm
\end{center} 
\vspace{1.0cm}

\vspace{6pt}
\begin{center}{\large \bf Abstract}

We examine in detail the semi-leptonic decay $\Lambda_b \to \Lambda_c \tau{\bar \nu}_{\tau}$, which may confirm previous hints, from the analogous $B$ decay, of a new physics beyond the standard model. First of all, starting from rather general assumptions, we predict the partial width of the decay. Then we analyze the effects of five possible new physics interactions, adopting in each case five different form factors.
In particular, for each term beyond the standard model, we find some constraints on the strength and phase of the coupling, which we combine with those found by other authors in analyzing the analogous semi-leptonic decays of $B$. Our analysis involves some dimensionless quantities, substantially independent of the form factor, but which, owing to the constraints, turn out to be strongly sensitive to the kind of non-standard interaction. We also introduce a criterion thanks to which one can discriminate among the various new physics terms: the left-handed current and the two-higgs-doublet model appear privileged, with a neat preference for the former interaction. Lastly, we suggest a differential observable that could, in principle, help to distinguish between the two cases.  

\end{center}

\vspace{10pt}

\centerline{PACS numbers: 13.30.Ce, 12.15.-y, 12.60.-i}

\newpage

\section{Introduction}
The high energy physicists have been looking for physics beyond the standard model (SM) for some decades. This research has recently received a new impulse from the Higgs discovery\cite{cms,atl} and from the data of the semi-leptonic decays  $B \to D^{(*)} \tau \nu_{\tau}$[3-8] and $B\to K^*\ell^+\ell^-$\cite{lhcb1,lhcb2}\footnote{$\ell$ denotes a light lepton, unless otherwise stated.}, which have exhibited strong tensions with the SM predictions[11-13]. Indeed, the SM entails lepton flavor universality (LFU), which seems to be contradicted by the measurements of the observables
\begin{equation}
R_{D^{(*)}} = \frac{{\cal B}(B\to D^{(*)} \tau \bar{\nu}_{\tau})}{{\cal B}(B\to D^{(*)} \ell \bar{\nu}_{\ell})} \ ~~~ {\mathrm and} \ ~~~
R_{K^*} = \frac{{\cal B}(B\to K^*\mu^+\mu^-)}{{\cal B}(B\to K^*e^+e^-)}. \label{ratb}
\end{equation}
It is important to notice that these quantities attenuate the biases related to the experimental efficiency, to the values of CKM matrix elements $V_{cb}$ and $V_{sb}$ and to the theoretical uncertainties of the form factors (FF); therefore they appear especially suitable for singling out new physics (NP) effects.

In the present article, we are mainly concerned with the experimental results of the $B \to D^{(*)} \tau \nu_{\tau}$ decays, about which some authors have performed model independent analyses[14-23], while other people have interpreted them in terms of specific NP models, like two-higgs-doublet[24-27] (2HD),  leptoquark[14,28-34] (LQ), left-right symmetric\cite{hv1,alt} (LR) or extra-dimension\cite{bsw} model. The anomaly has been connected to the leptonic $B$ and $B_c$ decays to $\tau \bar{\nu}_{\tau}$[24-27,35] and a new light has been cast on the muon anomalous magnetic moment\cite{chn,cai} (see also ref. 38). 
 
All that is a goad to further searches for confirmations of NP. In this sense, the $\Lambda_b$ decays to $\Lambda\ell^+\ell^-$\cite{dm,gu} and to $\Lambda_c\tau \nu_{\tau}$[41-47], as well as the decays $B_c\to J/\psi (\eta_c) \tau \nu_{\tau}$\cite{dub}, could give definitive confirmations of NP, in particular of LFU violation (LFUV); indeed, these presumably share the same basic processes as the two above mentioned $B$ decays. In the present paper, we consider the baryonic decay
\begin{equation}
\Lambda_b \to \Lambda_c \tau^- \bar{\nu}_{\tau},  \label{slh} 
\end{equation}
to which a previous letter\cite{dsa} has been dedicated. Here we give a more in-depth, model independent analysis of this decay, which we compare to the $\Lambda_b \to \Lambda_c \ell \nu_{\ell}$ one. Precisely, we limit ourselves to the spin-independent observables and analyze the NP dependence of suitable dimensionless ratios, among which, analogously to (\ref{ratb}),
\begin{equation}
R_{\Lambda_c} = \frac{{\cal B}(\Lambda_b \to \Lambda_c \tau \bar{\nu}_{\tau})}{{\cal B}(\Lambda_b\to \Lambda_c \ell
\bar{\nu}_{\ell})}. \label{ratl}
\end{equation}
To this end, we propose for the NP interaction five different dimension 6 operators, chosen according to the most frequently used models - typically, the above mentioned 2HD, LQ or LR - and similarly to other model independent analyses\cite{swd,dut,dt2}. The main differences with the previous studies consist of imposing more stringent constraints on the NP effects, also by taking into account some analyses of the semi-leptonic $B$ decays\cite{faj2,iv,tw}, and of introducing a particular criterion for discriminating among the different NP interactions. 
Moreover, in order to probe the FF dependence of our predictions, we consider five different alternatives. We find that, while the partial decay width depends rather strongly on the FF, the above mentioned ratios depend much more mildly on them. On the contrary, our prediction about $R_{\Lambda_c}$ is quite different from the one of the other authors[49-56]. However, as we shall see, there are reasons for assuming that the rest of the analysis is independent of this discrepancy.
As a last subject, we single out a differential observable which allows, in principle, to distinguish between two of the most likely NP interactions. 

Sect. 2 resumes our assumptions, including the above  mentioned criterion. In sect. 3, we deduce, in the covariant formalism, the general formulae for the matrix elements; we also introduce the various FF, for which we give a short review of previous contributions. In sect. 4, we sketch the expressions of the differential and partial widths of the decays of interest. In sect. 5, we show predictions of the partial decay widths, both according to the SM and to our assumptions about NP. Sect. 6 is devoted to illustrating the constraints on the various NP couplings.
Sect. 7 is dedicated to a discussion of our results, in light our criterion, and to a review of previous analyses. In sect. 8, we exhibit the predictions of the differential decay widths according to two different NP interactions, suggesting a new observable, sensitive to the differences between them. Lastly, some conclusions are presented in sect. 9  
 
\section{Assumptions}

We list here our assumptions, five in all. The first four are
shared by the other authors, whereas the fifth one is the above mentioned criterion.

1) The NP process entails LFUV, therefore it does not act on $\tau$ in the same way as on the light leptons. In a simplifying assumption, the NP does not concern at all the electron and the muon.

2) The basic process that gives rise to the NP in the semi-leptonic decays  (\ref{slh}) and $B \to D^{(*)} \tau \nu_{\tau}$ consists uniquely of $b \to c \tau \nu_{\tau}$ and does not involve any spectator partons.
This is a consequence of the short range of the would-be NP interaction, whose intermediate boson is estimated to have  a mass of order 1 $TeV$\cite{bln,frt,bdn,choh}. As we shall see, this has important consequences. 

3) Only one type of interaction $-$ scalar, vector, {\it etc.} $-$ is present in the effective lagrangian.
    
4) The double ratio 
\begin{equation}
R_{\Lambda_c}^{ratio} = R_{\Lambda_c}/R_{\Lambda_c}^{SM}
\end{equation}
depends only mildly on the FF. This assumption is supported by the analyses relative to the semi-leptonic $\Lambda_b$\cite{swd,lyz} and $B$[3-8,28,29] decays. In particular, according to refs. 28 and 58, it results
\begin{equation}
R_D^{ratio} = R_D^{exp}/R_D^{SM} = 1.30\pm0.17, \ ~~~~~ \ \ ~~~~~    R_{D^*}^{ratio} = R_{D^*}^{exp}/R_{D^*}^{SM} = 1.25\pm0.08, \label{rat-exp}
\end{equation}
quite compatible with each other. Further arguments will be exposed below.

5) Lastly, given the reliability of the SM at present energies, the NP term is only a perturbation of the known amplitude for the decay considered. Therefore, we privilege the interactions whose effective couplings are much smaller than the Fermi constant, $G = 1.166379 \cdot 10^{-5} GeV^{-2}$.  

Taking into account the more restrictive of the results (\ref{rat-exp}), the first four assumptions imply immediately that  
\begin{equation}
R_{\Lambda_c} = \xi \frac{\Gamma_{\tau}^{SM}}{\Gamma_{\ell}^{SM}}, \ ~~~~~ \ \ ~~~~~ \ \xi = 1.25 \pm 0.08. \label{rat-prd}
\end{equation} 
Here $\Gamma_{\tau(\ell)}$ is the partial width of the decay $\Lambda_b \to \Lambda_c \tau^- \bar{\nu}_{\tau} (\ell^- \bar{\nu}_{\ell})$; 
according to our  prediction, $\Gamma_{\tau}$ results to be
\begin{equation}
\Gamma_{\tau} = \xi \Gamma_{\tau}^{SM}. \label{gam-tau}
\end{equation}

\section{Matrix Element of the Decay}
\subsection{SM and NP Amplitudes}

We consider the matrix element for the decay $\Lambda_b \to \Lambda_c \ell{\bar \nu}_{\ell}$\footnote{In this section and in the next one, $\ell$ denotes either $\tau$ or a light lepton.}. To this end, we set, in quite a general way,
\begin{equation}
{\cal M} = V_{cb} \frac{G}{\sqrt{2}} (J^L_{\mu} j^{\mu}+ g_r {\cal I}). \label{matlm}
\end{equation}
Here ${\cal I}$ is the NP interaction and
\begin{equation} 
g_r = x e^{i\varphi}
\end{equation} 
the corresponding relative coupling\cite{swd}, with $x$ and $\varphi$ real, $x$ $>$ 0. We consider five types of effective dimension 6 operators, according to the most frequently used models:
\begin{equation}
{\cal I} = J^L_{\mu} j^{\mu}, ~~ J^R_{\mu} j^{\mu}, ~~ J^S j, ~~  J^P j, ~~ J^H j.
\end{equation}

Here
\begin{eqnarray}
j_{\mu} &=& {\bar u}_{\ell}\gamma_{\mu}(1-\gamma_5)v, \ ~~~~~ \ \ ~~~~~ \ \ ~~~~~ \ \ ~~ \ 
j = {\bar u}_{\ell}(1-\gamma_5)v,
\\
J^{L(R)}_{\mu} &=& \langle\Lambda_c|{\bar c}\gamma_{\mu}(1\mp\gamma_5)b |\Lambda_b\rangle,\ ~~~~~ \ \ ~~~~~ \ 
J^S = \langle\Lambda_c|{\bar c} b |\Lambda_b\rangle, 
\\
J^P &=& \langle\Lambda_c|{\bar c} \gamma_5 b |\Lambda_b\rangle, 
\ ~~~~~ \ \ ~~~~~ \ \ ~~~~~ \ ~~~ \ J^H = J^S - J^P 
\end{eqnarray}
and $u_{\ell}$ and $v$ are the four-spinors of the charged lepton and of the anti-neutrino respectively; lastly, $L$, $R$, $S$, $P$ and $H$ denote, respectively, left-handed vector, right-handed vector, scalar, pseudo-scalar and $S-P$-interaction. 

\subsection{Form Factors}

The most general expressions of the vector and axial hadronic currents are
\begin{eqnarray}
\langle\Lambda_c|{\bar c}\gamma_{\mu}b |\Lambda_b\rangle &=& 
\bar{u_f} V_{\mu}u_i = \bar{u_f} (f_1 \gamma_{\mu} + f_2 i\sigma_{\mu\nu} q^{\nu} + f_3 q_{\mu})u_i, \label{ffv} 
\\
\langle\Lambda_c|{\bar c}\gamma_{\mu}\gamma_5 b |\Lambda_b\rangle &=&
\bar{u_f} A_{\mu}\gamma_5 u_i = \bar{u_f} (g_1 \gamma_{\mu} + g_2 i\sigma_{\mu\nu} q^{\nu} + g_3 q_{\mu}) \gamma_5 u_i. \label{ffa}
\end{eqnarray}
Here the $f_i$ and the $g_i$ ($i$ = 1,2,3) are functions of $q^2$, $u_{i(f)}$ the four-spinor of the initial (final) baryon,
\begin{equation}
q = p_i-p_f = p_{\ell}+p
\end{equation}
and $p_{i(f)}$, $p_{\ell}$ and $p$ are, respectively, the four-momenta of the baryons, of the charged lepton and of the anti-neutrino. 

Using the equations of motion (eom), the operators $V_{\mu}$ and $A_{\mu}$, which appear in Eqs. (\ref{ffv}) and (\ref{ffa}), can be re-written as (see Appendix A) 
\begin{equation}
V_{\mu} = X_0 \gamma_{\mu} + f_2 P_{\mu} + f_3 q_{\mu},
\ ~~~~~ \
A_{\mu} = Y_0 \gamma_{\mu} + g_2 P_{\mu} + g_3 q_{\mu},
\end{equation}
where 
\begin{equation}
X_0 =  f_1-(m_i+m_f)f_2, ~~~~~ Y_0 = g_1+(m_i-m_f)g_2, ~~~~~ P = p_i+p_f 
\end{equation}
and $m_{i(f)}$ is the mass of the initial (final) baryon: $m_i$ = 5.619 $GeV$, $m_f$ = 2.286 $GeV$.

Moreover, as regards the (pseudo-) scalar current, still, the eom imply\cite{dsa}
\begin{equation}
J^S = \frac{q^{\mu}}{\delta m_Q}{\bar u}_f V_{\mu} u_i, ~~~~~ J^P = -\rho\frac{q^{\mu}}{\delta m_Q}{\bar u}_f A_{\mu} u_i, 
\end{equation}
with 
\begin{equation}
\delta m_Q = m_b-m_c, ~~~~~
\rho = \frac{m_b-m_c}{m_b+m_c} \sim 0.53, \label{2hdm}
\end{equation}
$m_b$ = 4.18 $GeV$ and $m_c$ = 1.28 $GeV$ being the masses of the $b$- and $c$-quark respectively.

\subsubsection{A Short Review}

Different techniques have been adopted for determining the FF for the decay (\ref{slh}): 

- lattice calculation\cite{dm1,dt2}, approximated by an analytical expression\cite{dut};
 
- quark models: constituent\cite{pv}, covariant\cite{guf}, diquark\cite{fgk} and heavy quark Isgur-Wise\cite{klw,lhcb4} (IW) model; 

- sum rules (SR), both in pole approximation\cite{dec,swd,lyz} and in full QCD\cite{azs}.

\subsubsection{Present Analysis}

The five different FF we use here are based on some approximations, generally accepted for the heavy quark transition $b\to c$\cite{dec}: 
\begin{equation}
f_1 = g_1, \ ~~~~~ \ f_2 = g_2 = A, \ ~~~~~ \ f_3 = g_3 = 0. \label{hff}
\end{equation}

In particular, the first FF is of the IW type\cite{klw} and the remaining four are based on the SR\cite{dec,swd,lyz}. The IW FF reads as
\begin{eqnarray}
f_1 (q^2) &=& \zeta_0 [\omega(q^2)] = 
1 -1.47[\omega(q^2)-1]+0.95[\omega(q^2)-1]^2, \label{iwff0}
\\ 
\omega(q^2) &=& \frac{m_i^2+m_f^2-q^2}{2m_im_f}; \ ~~~~ ~~~~ \ f_2 (q^2) = 0. \label{iwff}
\end{eqnarray}
Incidentally, it is worth noticing that this is quite compatible with the bounds determined by the recent analysis of $\Lambda_b \to \Lambda_c \mu \nu_{\mu}$ data\cite{lhcb4}.

The parametrizations of the SR FF are reported in Table 1. 

\begin{table*}
\begin{center}
\caption{The four different FF inferred from sum rules: $f_1$ is dimensionless, $f_2$ is expressed in $GeV^{-1}$ and $q^2$ in $GeV^2$.}
\begin{tabular}{|c|c|c|c|c|}
\hline\hline
$~~~~~~~~~~$&$~~~~SR1~~~~$&$~~~~SR2~~~~$&$~~~~SR3~~~~$&$~~~~SR4~~~~$ \\
\hline\hline
 $f_1(q^2)$ & 6.66/(20.27 - $q^2$) & 8.13/(22.50 - $q^2$) & 13.74/(26.68 - $q^2$) & 16.17/(29.12 - $q^2$) \\
 $f_2(q^2)$ & -0.21/(15.15 - $q^2$) & -0.22/(13.63 - $q^2$) & -0.41/(18.65 - $q^2$) & -0.45/(19.04 - $q^2$) \\
\end{tabular}
\label{tab:one}
\end{center}
\end{table*}

\section{Decay Width}
\subsection{Derivation of Basic Formulae}

The observables that we study in this paper are derived from 
\begin{equation}
d\Gamma = \frac{1}{2m_i} \sum|{\cal M}|^2 d\Phi. \label{ddw}
\end{equation}
Here $d\Phi$ is the phase space and the symbol $\sum$ denotes the average over the polarization of the initial baryon and the sum over the  polarizations of the final particles. We have 
\begin{equation}
\sum |{\cal M}|^2 = |V_{cb}|^2 \frac{G^2}{2} [T_{SM} + 2x
\Re(T_I e^{-i\varphi}) + x^2 T_N]. \label{modsq}
\end{equation}
Here $T_{SM}$ is the SM contribution,
\begin{equation}
T_{SM} = \sum H_{\mu\nu} \ell^{\mu\nu}, \ ~~~~~ \ H_{\mu\nu} = J^L_{\mu} J^{L*}_{\nu}, \ ~~~~~ \ \ell_{\mu\nu} = j_{\mu} j^*_{\nu}.
\end{equation}

As to the terms $T_I$ and $T_N$, they correspond, respectively, to the interference between the  SM and the NP amplitude and to the modulus square of the NP amplitude. Specifically, we have
\begin{eqnarray}
T_I^L &=& T_N^L = T_{SM}, \ ~~~ \
T_I^R = \sum J^L_{\mu} J^{R*}_{\nu}\ell^{\mu\nu}, 
\ ~~~ \ 
T_N^R = \sum J^R_{\mu} J^{R*}_{\nu}\ell^{\mu\nu},
\\
T_I^{S(P)} &=& \sum J^L_{\mu} J^{S(P)*} j^{\mu}j^*, 
\ ~~~ \ \ ~~~ \
T_N^{S(P)} = \sum J^{S(P)} J^{S(P)*} jj^*,
\\  
T_I^H &=& \sum J^L_{\mu} J^{H*} j^{\mu}j^*, 
\ ~~~ \ \ ~~~ \ \ ~~~ \
T_N^H = \sum J^H J^{H*} jj^*,	 
\end{eqnarray}

the upper indices denoting the various NP interactions. 

In the present paper we are not concerned with spin, therefore we consider an unpolarized initial baryon. A standard calculation in the covariant formalism leads to 
\begin{align}
T_{SM} &= 2^5 \{(X_0+Y_0)^2h_1 + (X_0-Y_0)^2h_2 + (Y_0^2-X_0^2)h_3 \ ~~~~~  ~~~~~ \  \nonumber
\\
\ ~~~~~ \ &+ A[m_f(X_0+Y_0){\cal L}_i+ m_i(X_0-Y_0){\cal L}_f]+A^2p_f\cdot p_i ~ {\cal L}_P\}, \label{smc}
\end{align}
where $A$ is defined by the second Eq. (\ref{hff}) and 
\begin{eqnarray}
h_1 &=& p_f\cdot p_{\ell} ~ p_i \cdot p, \ ~~~~~ \ h_2 = p_f \cdot p 
 ~ p_i\cdot p_{\ell}, \ ~~~~~ \  h_3 = m_i m_f ~ p \cdot p_{\ell},
\\ 
{\cal L}_{i(f)} &=& p_{i(f)}\cdot p_{\ell} ~ P\cdot p + p_{i(f)}\cdot p ~ P\cdot p_{\ell}-p_{i(f)}\cdot P ~ p\cdot p_{\ell},
\\
{\cal L}_P &=& 2p_{\ell}\cdot P ~ p\cdot P - P^2 ~ p \cdot p_{\ell}.	 
\end{eqnarray}

As regards the remaining terms, one has
\begin{eqnarray}
T_I^R &=& 2^6 \{(X_0^2-Y_0^2)(k_1+k_2) - (X_0^2+Y_0^2)k_3 \nonumber
\\
&+& A[m_f(X_0+Y_0){\cal L}_i+ m_i(X_0-Y_0){\cal L}_f] 
+ A^2 m_f m_i{\cal L}_P\},
\\
T_N^R &=& 2^5 \{(X_0-Y_0)^2k_1 +(X_0+Y_0)^2k_2 + (Y_0^2-X_0^2)k_3
\nonumber
\\
&+& A[m_f(X_0+Y_0){\cal L}_i+ m_i(X_0-Y_0){\cal L}_f] \nonumber
+ A^2p_f\cdot p_i{\cal L}_P\},
\\
T_I^S &=& 2^5 \frac{m_l}{\delta m_Q} [X_0^2(k_1+k_2)+A X_0(k_3+k_4)+A^2 
p\cdot P ~ q\cdot P ~ k_+],
\\
T_N^S &=& 2^4 \frac{p_l\cdot p}{(\delta m_Q)^2}[X_0^2(k_5+k_6)+A X_0(k_7+k_8)+A^2 (q\cdot P)^2 ~ k_+],
\\
T_I^P &=& 2^5\frac{m_l}{\delta m_Q} [Y_0^2(-k_1+k_2)+A Y_0(k_3-k_4)-A^2 
p\cdot P ~ q\cdot P ~ k_-],
\\
T_N^P &=& 2^4\frac{p_l\cdot p}{(\delta m_Q)^2}[Y_0^2(k_5-k_6)+
A Y_0(-k_7+k_8) + A^2 (q\cdot P)^2 ~ k_-],
\\
T_I^H &=& T_I^S + \rho T_I^P, \ ~~~~~ \ T_N^H = T_N^S + \rho^2 T_N^P.   
\end{eqnarray}
Here 
\begin{eqnarray}
k_1 &=& p\cdot p_f ~ q\cdot p_i+ p\cdot p_i ~ q\cdot p_f-p\cdot q  ~ p_f\cdot p_i, \ ~~~ \ ~~~ \  k_2 = m_i m_f ~ p\cdot q,
\\
k_3 &=& m_i(p\cdot p_f ~ q\cdot P + p\cdot P ~ q\cdot p_f), \ ~~~ \  k_4 = m_f(p\cdot p_i ~ q\cdot P + q\cdot p_i ~ p\cdot P),
\\ 
k_5 &=& 2q^2 ~ p_f\cdot q ~ p_i\cdot q ~ p_i\cdot p_f, \ ~~~ \ \ ~~~ \ \ ~~~~ \ k_6 = m_i m_f ~ q^2, 
\\   
k_7 &=& m_i ~ p_f\cdot q ~ P\cdot q, \ ~~~ \ \ ~~~~~ \ \ ~~~ \ \ ~~~~~ \ k_8 = m_f ~ p_i\cdot q ~ P\cdot q, 
\\
k_+ &=& p_i\cdot p_f + m_i m_f, \ ~~~ \ \ ~~~ \ \ ~~~ \ \ ~~~ \ \ ~ \ k_- = p_i\cdot p_f - m_i m_f. 
\end{eqnarray}

\subsection{Differential and Partial Decay Width}

The integration over the phase space is suitably performed by fixing a reference frame at rest with 
respect to $\Lambda_b$; to this end, it is also worth recalling the relation of $q^2$ to the energy 
$E_f$ of the final baryon in that frame:
\begin{equation}
q^2 = m_i^2+m_f^2-2 m_iE_f.  \label{q2}
\end{equation}

After integrating Eq. (\ref{ddw}) over the angular variables, the differential decay width reads as\cite{dsa}
\begin{equation}
\frac{d\Gamma_{\ell}}{dq^2} = \frac{1}{2^7 \pi^3 m_i^2} \int_{E_{\ell}^-}^{E_{\ell}^+} dE_{\ell}\sum|{\cal M}|^2. \label{ddw1}
\end{equation}
Here $E_{\ell}$ is the energy of the charged lepton 
in the above mentioned frame and
\begin{eqnarray}
E_{\ell}^{\pm}&=& \frac{b\pm\sqrt{\Delta}}{2q^2},  \ ~~~~~ \ \ ~~~~~ \ 
\Delta = b^2+4q^2c, \label{zrs}
\\
b &=& 2m_iE_f^2 - (2m_i^2+M^2)E_f+M^2m_i, \ ~~~~~ \ \ ~~~~~ \ 
\\
c &=& -(m_i^2+m^2_{\ell})E_f^2+m_iM^2E_f+m_f^2 m^2_{\ell}-\frac{1}{4}M^4, \label{coef2} \ ~~~~~ \  
\\
M^2 &=& m_i^2+m_f^2+m_{\ell}^2; \ ~~~~~ \ \ ~~~~~ \ \ ~~~~~ \ \ ~~~~~ \ \label{coef3}
\end{eqnarray}
moreover, $m_{\ell}$ = 0.106 $GeV$ for $\ell$ = $\mu$ and 1.777 $GeV$ for $\ell$ = $\tau$.

The partial decay width is obtained by integrating Eq. (\ref{ddw1}) over $q^2$:
\begin{equation}
\Gamma_{\ell} = \int_{q^2_-}^{q^2_+}dq^2\frac{d\Gamma}{dq^2}. \label{pdw1}
\end{equation}
Here the limits $q^2_{\pm}$ are related, through Eq. (\ref{q2}), respectively to $E_f$ = $m_f$ and $E_f$ = $E_f^m$, with 
\begin{equation}
E_f^m = \sqrt{m_f^2+ p_m^2}, \ ~~~~~ \ \ ~~~~~ \  p_m = \frac{1}{2}(m_i-m_{\ell}-\frac{m_f^2}{m_i-m_{\ell}}).\label{lmf}
\end{equation}

For later convenience, we re-write the partial decay width, Eq. (\ref{pdw1}), as 
\begin{equation}
\Gamma_{\ell} = \Gamma_{\ell}^{SM}+2 x cos\varphi \Gamma_{\ell}^I + x^2 \Gamma_{\ell}^N. \label{meq}
\end{equation}
Here, taking account of Eqs. (\ref{ddw1}) and (\ref{modsq}), we have  
\begin{equation}
\Gamma_{\ell}^{SM} = \frac{|V_{cb}|^2}{2^7 \pi^3 m_i^2} \frac{G^2}{2} 
\int_{q^2_-}^{q^2_+}dq^2 \int_{E_{\ell}^-}^{E_{\ell}^+} dE_{\ell} T_{SM},  \label{meq1}
\end{equation}
similar expressions holding for $\Gamma_{\ell}^I$ and $\Gamma_{\ell}^N$, with $T_I$ and $T_N$ in place of $T_{SM}$. A check of the formulae used is given in Appendix B, where, in particular, the expression of $\Gamma_{\ell}^{SM}$ is compared with the well-known formula of the muon decay. 
 
\section{Predictions of Partial Decay Widths}

\begin{table*}
\begin{center}
\caption{$\Gamma_{\mu}^{SM}$, $\Gamma_{\tau}^{SM}$ (in $\mu eV$) and the ratio $R_{\Lambda_c}^{SM}$ =  $\Gamma_{\tau}^{SM}/\Gamma_{\mu}^{SM}$, for the five different FF considered.}
\begin{tabular}{|c|c|c|c|}
\hline\hline
$~~~~~FF~~~$&$~~~~\Gamma_{\mu}^{SM}~~~~$&$~~~~\Gamma_{\tau}^{SM}~~~~$&$~~~~R_{\Lambda_c}^{SM}~~~~$ \\
\hline\hline
 IW   & 31.6 & 5.63 & 0.18 \\
 SR1  & 10.8 & 1.95 & 0.18 \\
 SR2  & 11.5 & 1.80 & 0.16 \\ 
 SR3  & 22.1 & 3.40 & 0.15 \\
 SR4  & 24.5 & 3.61 & 0.15 \\
\end{tabular}
\label{tab:two}
\end{center}
\end{table*}

Table 2 shows the values of $\Gamma_{\mu}^{SM}$ and $\Gamma_{\tau}^{SM}$, calculated by means of Eq. (\ref{meq1}), and the ratio $R_{\Lambda_c}^{SM}$ =  $\Gamma_{\tau}^{SM}/\Gamma_{\mu}^{SM}$, for the five different FF considered in the article.
It can be seen that the SM results of the partial widths depend strongly on the FF. In particular, as regards $\Gamma_{\mu}^{SM}$, the IW FF gives the best approximation of the experimental value, {\it i. e.}\cite{pdg},
\begin{equation} 
\Gamma_{\ell}^{exp} = (29.5_{-11.4}^{+14.5}) \mu eV. \label{exp}
\end{equation} 
Our result agrees also with the numerical value given in ref. 53.
 Instead, two of the SR FF differ from this value by more than one standard deviation and probably they need an overall normalization factor. However, we consider in the present article mainly ratios between dimensional quantities, which appear to be barely dependent on the FF. A first example is offered by the ratio $R^{SM}_{\Lambda_c}$, listed in the last column of Table 2. 

This table and Eq. (\ref{rat-prd}) entail a prediction for $R_{\Lambda_c}$. Indeed, averaging over the five values yields   
\begin{equation}
{\bar R}_{\Lambda_c}^{SM} = 0.164\pm 0.006, \ ~~~~~ \ {\bar R}_{\Lambda_c} = 0.205 \pm 0.013\pm 0.008.\label{rat-prd1}
\end{equation}
Here the former ratio is only affected by the systematic error caused by the FF uncertainty, while for the latter also the statistical one (0.013) has to be accounted for. The smallness of the theoretical error confirms assumption 4). A particular attention deserves the IW FF. First of all, it allows to check immediately our formula (\ref{meq1}) against the expression of the well-known muon decay width, as shown in Appendix B. Secondly, it yields, for $R_{\Lambda_c}^{SM}$ and for the other dimensionless quantities considered in our article, results that are similar to those obtained with the SR FF, although structurally different.

On the contrary, our result for $R_{\Lambda_c}^{SM}$ is considerably smaller than those given by the other authors. Indeed, such values span from 0.26\cite{lyz} to 0.38\cite{pv}, being concentrated, in recent years, between 0.31 and 0.34\cite{gu1,dm1,dut,fgk,swd,dt2,azs}.
Refs. 52 and 56 give more complete reviews of these results.
In any case, the analysis exposed in the following sections is presumably independent of such a discrepancy, as it is based on the ratios $\chi$ and $r_{\pm}$ (Eqs. (\ref{cc1}) and (\ref{rr1}) respectively), which depend exclusively on the decay (\ref{slh}).

\section{Couplings of the Various NP Interactions}

\subsection{Argand Diagrams for the NP Couplings}

\begin{table*}
\begin{center}
\caption{Values of $\Gamma_{\tau}^I$ and $\Gamma_{\tau}^N$ (in $\mu eV$) for $S$, $P$ and $R$-interactions and for the five different FF}
\begin{tabular}{|c|c|c|c|c|c|c|}
\hline\hline
$~~~~~FF~~~$&$~~~~\Gamma_{\tau}^{I,S}~~~~$&$~~~~\Gamma_{\tau}^{I,P}~~~~$&$~~~~\Gamma_{\tau}^{I,R}~~~~$&$~~~~\Gamma_{\tau}^{N,S}~~~~$&$~~~~\Gamma_{\tau}^{N,P}~~~~$&$~~~~\Gamma_{\tau}^{N,R}~~~~$ \\
\hline\hline
 IW~  & 1.28 &  0.26 & -3.32 & 2.26 & 0.44 & 5.63 \\
 SR1  & 0.58 &  0.12 & -0.67 & 1.03 & 0.19 & 1.95 \\
 SR2  & 0.60 &  0.12 & -0.39 & 1.06 & 0.20 & 1.80 \\ 
 SR3  & 0.99 &  0.20 & -1.21 & 1.75 & 0.34 & 3.40 \\
 SR4  & 1.05 &  0.22 & -1.25 & 1.86 & 0.36 & 3.61 \\
\end{tabular}
\label{tab:three}
\end{center}
\end{table*}

Table 3 provides the values of $\Gamma_{\tau}^I$ and $\Gamma_{\tau}^N$ for the $S$-, $P$- and $R$-interaction, calculated by Eq. (\ref{meq}) together with the equations analogous to (\ref{meq1}). The parameters corresponding to the $H$-interaction can be deduced from the following linear combinations: 
\begin{equation}
\Gamma_{\tau}^{I,H} = \Gamma_{\tau}^{I,S} - \rho \Gamma_{\tau}^{I,P},
~~~~~ \ ~~~~~ \Gamma_{\tau}^{N,H} = \Gamma_{\tau}^{N,S} + \rho^2 \Gamma_{\tau}^{N,P}. \label{twhg}
\end{equation} 
As regards the $L$-interaction, we have the re-scaling
\begin{equation}
|1+x_L e^{i\varphi}|^2 = \xi, \label{lhd}
\end{equation}
independent of the FF. 

Eq. (\ref{meq}) yields, together with Eq. (\ref{gam-tau}), a relation between $x$ and $\varphi$. Taking account of the statistical and systematic errors, the allowed region consists of a circular crown in the Argand plane of the coupling $g_r$, centered at
\begin{equation}
g_c \equiv (\chi, 0), \ ~~~~ \ ~~~~ \ \chi = -\frac{\Gamma_{\tau}^I}{\Gamma_{\tau}^N}  \label{cc1}
\end{equation}
and with radii
\begin{equation}
r_{\pm} = \frac{\sqrt{\Delta_{r\pm}}}{\Gamma_{\tau}^N}, ~~~~ \ ~~~~ \ 
\Delta_{r\pm} = (\Gamma_{\tau}^I)^2 + (\Gamma_{\tau\pm} -\Gamma_{\tau}^{SM})\Gamma_{\tau}^N;  \label{rr1}
\end{equation}
here $\Gamma_{\tau\pm}$ takes into account the statistical error of $\xi_{\pm}$, Eq. (\ref{rat-prd}), and the systematic one, related to the FF. Exceptionally, the latter is absent for $L$-interaction, as Eq. (\ref{lhd}) entails, independent of the FF, 
\begin{equation}
g_c \equiv (-1, 0), ~~~~ r_{\pm} = \sqrt{\xi_{\pm}}.    \label{crr}
\end{equation}

The mean values and the statistical and systematic errors of the radii and the coordinates of centers of the Argand diagrams are listed in Table 4. Again, we remark the small theoretical errors of the parameters, which reflect the mild FF dependence.

\subsection{Remarks}

Two remarks are in order for the case of $\varphi$ = $\pm\pi/2$, where the interference between the SM amplitude and the NP one vanishes. 

- Firstly, note that comparing the results of Table 2 and of Table 3 yields
\begin{equation}
\Gamma_{\tau}^{N,R} = \Gamma_{\tau}^{SM};       \label{rel23}
\end{equation}
this is a consequence of the integration of Eq. (\ref{ddw}) over the phase space, which washes out the interference term between the vector and the axial current. Therefore we have, again independent of the FF, 
\begin{equation}
x_R(\pm\pi/2) =  x_L(\pm\pi/2) = 0.50\pm 0.04.                 \label{rel33}
\end{equation}

- Secondly, if one considers the possibility of decays $\Lambda_b \to \Lambda_c \tau^- \bar{\nu}_{\ell}$, with $\ell$ = $e$, $\mu$, $\tau$\cite{tw}, the coupling strength for $\ell$ = $\mu$ and $e$ can be inferred just for $\varphi$ = $\pm\pi/2$.

\begin{table*} 
\begin{center}
\caption{The mean values of the radii and of the centers of the Argand diagrams for the relative couplings $g_r$. $\bar{r}$ is affected both by a statistical and a systematic error, the former and the latter one respectively.}
\begin{tabular}{|c|c|c|c|c|c|}
\hline\hline
$~~~~~~~~$&$~~~~S~~~~$&$~~~~P~~~~$&$~~~~H~~~~$&$~~~~L~~~~$&$~~~~R~~~~$ 
\\
\hline\hline
 $\bar{r}$ & 0.90$\pm$0.04$\pm$0.02 & 
3.21$\pm$0.20$\pm$0.08 & 0.83$\pm$0.04$\pm$0.02 & 1.12$\pm$0.02 & 0.65$\pm$0.03$\pm$0.04 \\
 $g_c$ & (-0.56, 0) & (-1.12, 0) & (-0.48, 0) & (-1.0, 0) & (0.37$\pm$0.10, 0) \\
\end{tabular}
\label{tab:four}
\end{center}
\end{table*}

\subsection{Relative Strengths of the NP Interactions}

In order to compare the strengths of the various NP interactions, we may, for example, calculate their minimal values. These occur at $\varphi$ = 0, except for the $R$-interaction, for which one has to set $\varphi$ = $\pi$. This singular behavior is due to the negative value of $\Gamma_{\tau}^{I,R}$ (see Table 3), which induces, through Eq. (\ref{cc1}), a real positive value of $\chi$, and to the positivity of 
$x_{min} =  r-|\chi|$, which follows from Eqs. (\ref{cc1}) and (\ref{rr1}). As we shall see in a moment, this anomaly is connected to a strong limitation on the phase $\varphi$. The values of $x_{min}$ $-$ once more barely FF dependent $-$ are listed in Table 5. 

\subsection{Constraints from $B\to D^{(*)} \tau {\bar \nu}_{\tau}$ Decays}

Now we exhibit the phase limitations implied by our analysis, when combined with the analogous ones performed on the semi-leptonic $B$ decays\cite{ faj2,dt1,tw,iv,ddt}, especially the most recent ones\cite{iv}.

 - As regards the $L$-interaction, the agreement with all of the previous papers is trivial, because the NP term just re-scales the SM interaction.
                     
- As shown before, the minimum value of $x$ for the $R$-interaction occurs in correspondence of $\varphi$ = $\pi$. This property is shared by the $B \to D^* \tau \nu_{\tau}$ decay, whereas the $B \to D \tau \nu_{\tau}$ decay indicates that the minimum occurs at $\varphi$ = 0\cite{tw}. Therefore the allowed region for the coupling amounts to the intersection between two circular crowns whose centers are considerably far from each other, which strongly restricts the range of values of the phase; precisely, the previous analyses provide a narrow nearby of $\varphi$ = $\pm\pi/2$\cite{faj2,tw,iv}. But, as shown, in this case one has $x$ = 0.50 $\pm$ 0.04.

- Also the $H$-interaction exhibits strong limitations on its phase. Indeed, we have to take into account the results of Table 4, together with those by refs. 19, that is, $g_c$ = (-0.76, 0.0), $r$ = 1.03 $\pm$ 0.25\footnote{Ivanov {\it et al.}\cite{iv}, private communication.}, together with a bound on the phase\cite{tw,iv}. Then, the allowed region of the Argand plane amounts to two very small intervals around 
\begin{equation}
\varphi = \pm 2.18 ~ rad,       \label{rel47}
\end{equation}
as illustrated in Fig. 1. 

- Lastly, the $S+P$-interaction is excluded by recent analyses\cite{iv}; similarly, the tensor interaction does not find an appreciable room\cite{tw,iv}.

                          %%%%%%%%%%%%%%%%%%%%

%   FIGURE 1

\begin{figure}
\centering
\includegraphics[width=0.70\textwidth] {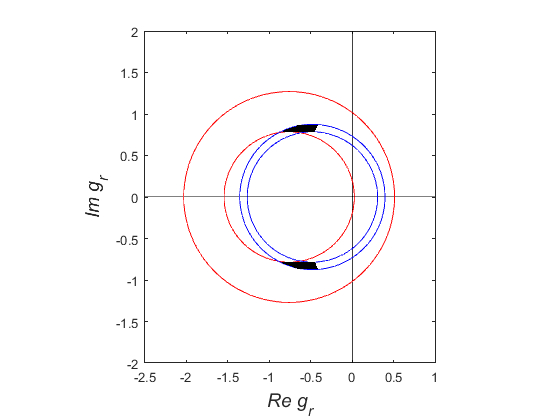}
\caption{$H$-interaction: the Argand diagram for the relative coupling $g_r$. The thinner of the two circular crowns is inferred from our analysis, the other one from the second ref. 19, where also bounds on the phase have been established. The dark regions correspond to the range of the allowed values of $g_r$.} 
\end{figure}

\section{Discussion}

First of all, we draw some consequences of our assumption 5) from the bounds just discussed. Secondly, we review and comment some of the previous analyses.

\subsection{Analyzing the Results}

- The $P$-interaction demands a quite large coupling ($x$ $>$ 2), in order to compensate the smallness of the matrix element of the corresponding operator between the initial and final state. This appears unrealistic, also in view of the considerations by Datta {\it et al.}\cite{dt2}, who discard this interaction when compared with the data of the decay $B_c \to \tau \bar{\nu}_{\tau}$.

- As a consequence, for $x$ $\leq$ 1, the $H$-interaction ($H$ = $S-P$) behaves quite similarly to the $S$ one, as can be seen from Tables 4 and 5. Moreover, as shown before, when combined with the previous ones, our analysis imposes strong limits on the phase, which entails a relative strength that is considerably greater than the minimal value. Indeed, in order to determine $x$, we set, at the left-hand side of Eq. (\ref{meq}), $\Gamma_{\ell}$ = $\Gamma_{\tau}$ = $\xi \Gamma_{\tau}^{SM}$ and fix $\varphi$ according to Eq. (\ref{rel47}). The smaller root of this equation yields
\begin{eqnarray} 
x_0 &=& 1.18\pm 0.12, ~~~~ x_1 = 1.09\pm 0.10, ~~~~ x_2 = 1.05\pm 0.09, \nonumber \\
x_3 &=& 1.09\pm 0.10, ~~~~ x_4 = 1.09\pm 0.10, ~~~~ \ ~~~~ \ ~~~~ \label{xh}
\end{eqnarray} 
where $x_0$ corresponds to the IW FF, the remaining $x_i$ to the SR FF.
Apart from the scarce agreement with our assumption 5), we observe that the 2HD model, included in the $H$-interaction, presents difficulties in explaining the anomaly[24-27], despite the fact that its coupling depends on the flavor, as required by LFUV.

- Similarly, the $R$-interaction $-$ implemented by a specific model\cite{bb} $-$ is affected, as seen, by strict limitations on the phase.

- On the contrary, as regards the $L$-interaction, any value of $\varphi$ is admitted by the analyses. This entails the possibility of a small ($\sim$ 0.12) value of the relative strength It is in qualitative agreement with a possible solution to the anomaly observed in the $B\to K^*\ell^+\ell^-$ decay\cite{gls,choh}, for which a very small relative strength is required. Moreover, this interaction is favored in the optics of MFV\cite{frt}, since it does not imply a CP violation phase out of the CKM scheme, as opposed to the cases of $H$- and $R$-interactions.

Incidentally, estimating the mass of the NP intermediate boson to be  about 10 times that of the usual intermediate vector boson $W$, this implies that the NP coupling  in the $H$-, $R$- and $L$-interaction is, respectively,  $\sim$ 10, 7 and 3.5 times greater than the electroweak coupling constant. 

To conclude our analysis, we remark that the $L$- and $H$-interactions recur in the most common models used to explain NP effects of the semi-leptonic decay and might be compatible with the anomaly seen in the $B\to K^*\ell^+\ell^-$ decay\cite{dln,crv,cai,choh,bht2}. According to
assumption 5), the former interaction appears favored. However, as we shall see in the following subsection, alternative analyses lead to different conclusions. Therefore, measurements for discriminating among different NP interactions are suitable, as we shall exhibit in the next section.

\subsection{Previous Analyses}

\subsubsection{$\Lambda_b \to \Lambda_c \tau{\bar \nu}_{\tau}$}

Two analyses\cite{swd, dt2} are quite similar to the present one, they also show Argand diagrams for the NP couplings. 
Shivashankara {\it et al.}\cite{swd} take into account the constraints that derive from $R_{D^{(*)}}$ and remark that the effects produced by the $P$-interaction are larger than those caused by the $S$-one. 
Datta {\it et al.}\cite{dt2} fix the NP couplings so that $R_{\Lambda_c}^{ratio}$ = $R_{D^{(*)}}^{ratio}$ within 3 standard deviations. Their condition is similar to ours, but less restrictive, therefore they get less severe bounds on phases and strengths of the couplings. 

Dutta\cite{dut} assumes that $R_{\Lambda_c}$ = $R_{D^{(*)}}$
within 3 standard deviations and considers two possible scenarios, either 
a mixing of $L$- and $R$-, or of $H$- and $S+P$-interactions. In the former case, he finds that only the $L$- or the purely vector interaction are possible. On the contrary, either the $H$- or the $S$-interaction survives the latter scenario, with more restrictions on the parameter space. This is not in contradiction with our results.

Li {\it et al.}\cite{lyz} analyze the decay in the framework of the leptoquark model, taking account of the $B\to \tau\nu$ decay. They examine either the vector or the scalar case, finding more restrictions 
for the latter alternative. 

\subsubsection{$B \to D^{(*)}\tau{\bar \nu}_{\tau}$}

We signal here two analyses of the $B$-decay, alternative to those considered above, which lead to different conclusions about the NP interaction.

S. Bhattacharya {\it et al.}\cite{bht} fit the FF to the data of the $B \to D^{(*)}\tau{\bar \nu}_{\tau}$ decay using only the SM term; then they compare such FF with those available from $B \to D^{(*)}\ell{\bar \nu}_{\ell}$ data, finding a disagreement only as regards the axial current. To choose among the various NP operators, they use information-theoretic approaches and goodness-of-fit tests for cross validation,
indicating the $R$-interaction as the best one.

Celis {\it et al.}\cite{clj} consider it difficult to explain LFUV with $L$- and $R$-interactions and perform a comprehensive analysis of the scalar contributions in $b\to c \tau \nu_{\tau}$ transitions. The authors examine various observables, like $R_{D^{(*)}}$, the $q^2$ differential distributions of $B \to D^{(*)}\tau{\bar \nu}_{\tau}$ and the $\tau$ polarization in $B \to D^*\tau{\bar \nu}_{\tau}$ and the $B_c$ lifetime.
They find that, in the framework of scalar NP, the discrepancy with the SM can be explained by a mixing of $H$- and $S+P$-interaction, with a slight tension for $R_{D^*}$. 
	 
\section{Alternative Observables for New Physics}
 
\subsection{Previous Proposals}

In order to discriminate among the possible NP interactions, various observables have been proposed for the semi-leptonic $\Lambda_b$ decays.
We recall especially the $\tau$ or $\Lambda_c$ polarization\cite{gu1},
the forward-backward asymmetry on the lepton side\cite{gu1,dt2} 
and the differential observable\cite{swd,dt2}
\begin{equation}
B_{\Lambda_c}(q^2) = \frac{d\Gamma_{\tau}}{dq^2}/\frac{d\Gamma_{\ell}}{dq^2}, \label{drb} 
\end{equation}
where $d\Gamma_{\tau(\ell)}/dq^2$ is the differential width of the semi-leptonic $\Lambda_b$ decay, with the $\tau$- $(\ell)$-lepton in the final state.
 
As regards the $B$ semi-leptonic decays, some asymmetry\cite{cg,ddt} and the polarization of one of the final products\cite{faj,lee2,tw,iv,chn,kum}, especially its T-odd component\cite{iv}, have been suggested. 

In this connection, we remark that a a $T$-odd observable could help to reveal a non-trivial phase $\varphi$, which seems to occur in the cases of $H$- and $R$-interactions.

\subsection{A New Suggestion}

As an alternative to the observables just exposed above, we propose the following one:
\begin{equation}
\Delta r(q^2) = \frac{B_{\Lambda_c}(q^2)}{B^{SM}_{\Lambda_c}(q^2)}-1 = \frac{d\Gamma_{\tau}}{dq^2}/(\frac{d\Gamma_{\tau}}{dq^2})_{SM}-1. 
\label{drq} 
\end{equation}  
Fig. 2 shows the behavior of this quantity in the case of the $H$-interaction, assuming, as found before, $\varphi$ = $\pm2.18$ $rad$ and the strengths (\ref{xh}) for the different FF. Once more, it does not depend so dramatically on the FF.

As regards the $L$-interaction, one has, independent of the FF, 
\begin{equation}
\Delta r(q^2) = 0.25 \pm 0.04 \label{lrc}
\end{equation}
for any $\varphi$. This is equal to the distribution (\ref{drq}) for the $R$-interaction at $\varphi$ = $\pm\pi/2$.
\begin{table*}
\begin{center}
\caption{Minimal values of the relative strength $x$ for the various interactions and for the different FF.}

\begin{tabular}{|c|c|c|c|c|c|}
\hline\hline
$~~~~~FF~~~$&$~~~~x_S~~$&$~~~x_P~~~~$&$~~~~x_H~~~~$&$~~~~x_L~~~~$&$~~~~x_R~~~~$ \\
\hline\hline
 IW~  & 0.41$\pm$0.10 & 2.44$\pm$0.51 & 0.43$\pm$0.10 & 0.12$\pm$0.04 &     0.18$\pm$0.05 \\
 SR1  & 0.33$\pm$0.09 & 2.08$\pm$0.45 & 0.35$\pm$0.09 & 0.12$\pm$0.04 &     0.26$\pm$0.07 \\
 SR2  & 0.30$\pm$0.08 & 1.91$\pm$0.42 & 0.32$\pm$0.08 & 0.12$\pm$0.04 &     0.33$\pm$0.07 \\ 
 SR3  & 0.33$\pm$0.09 & 2.08$\pm$0.45 & 0.35$\pm$0.09 & 0.12$\pm$0.04 &     0.26$\pm$0.07 \\
 SR4  & 0.33$\pm$0.09 & 2.06$\pm$0.45 & 0.35$\pm$0.09 & 0.12$\pm$0.04 &     0.26$\pm$0.07 \\
\end{tabular}
\label{tab:five}
\end{center}
\end{table*}

\section{Conclusions}

Let us stress the most relevant points of our paper.

A) As already observed in sect. 5, the results concerning the partial widths depend rather strongly on the FF. On the contrary, the dimensionless parameters $r$, $\chi$ and $x$, as well as the observables $R_{\Lambda_c}$, $R^{ratio}_{\Lambda_c}$ and $\Delta r(q^2)$, exhibit, similarly to ref. 41, a mild FF dependence, contained within $\sim$ $2-3\%$. Actually, such uncertainties vanish at all if the $L$-interaction is assumed. Yet, our prediction on $R_{\Lambda_c}$ differs considerably from those by other authors.

B) We have done some assumptions, generally shared by the other authors, 
furthermore we have adopted a particular criterion for choosing the type of NP interaction. We have also taken into account the analyses of the $B$ semi-leptonic decays and the most commonly used models. On this basis, our calculations indicate that the most likely NP interactions are the $L$- and $H$-one. But the former interaction appears simpler and more natural. 

C) Our conclusions about the NP term are in contrast with those by other authors. At this point the measurements of alternative observables, like polarization or other asymmetries, is determinant. In particular, we have proposed a differential observable which could allow to discriminate between the two NP interactions mentioned at point B). 

                    %%%% FIGURE 2

\begin{figure}
\centering
\includegraphics[width=0.70\textwidth] {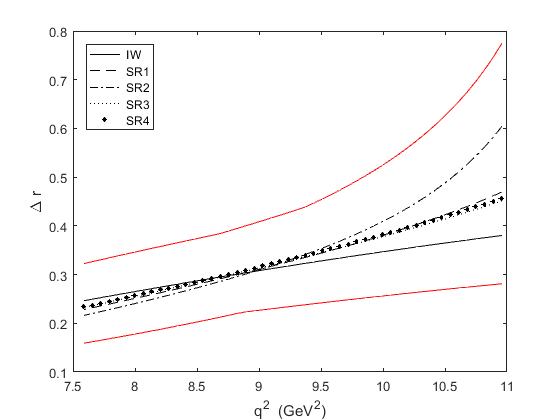}
\caption{The observable $\Delta r$, Eq. (\ref{drq}), as a function of $q^2$, with $\varphi$ = $\pm$ 2.18 $rad$; see Eqs. (\ref{xh}) for the corresponding values of $x$. The upper and lower curve delimit the allowed band.} 
\end{figure}

\vskip 0.50cm

                 \centerline{\bf Acknowledgments}

The authors are thankful to their colleagues Fajfer {\it et al.}\cite{faj,faj2} and Ivanov {\it et al.}\cite{iv} for helpful communications and suggestions.

\newpage

\setcounter{equation}{0}
 \renewcommand\theequation{A. \arabic{equation}}

 \appendix{\large \bf Appendix A}
\vskip 0.30cm

Here we show that the operators 
\begin{equation}
V_{\mu} = f_1 \gamma_{\mu} + f_2 i\sigma_{\mu\nu} q^{\nu} + f_3 q_{\mu} \label{op1}
\end{equation}
and 
\begin{equation}
A_{\mu} = (g_1 \gamma_{\mu} + g_2 i\sigma_{\mu\nu} q^{\nu} + g_3 q_{\mu}) \gamma_5, \label{op2}
\end{equation}
when inserted between the initial and the final baryon state, can be re-written as
\begin{eqnarray}
V_{\mu} &=& [f_1-(m_i+m_f)f_2]\gamma_{\mu} + f_2 P_{\mu} + f_3 q_{\mu},  \label{eqv}
\\
A_{\mu} &=& \{[g_1+(m_i-m_f)g_2]\gamma_{\mu} + g_2 P_{\mu} + g_3 q_{\mu}\}\gamma_5, \label{eqa}
\end{eqnarray}
thanks to the equations of motion (eom). Here 
\begin{equation}
q = p_i-p_f, \ ~~~~~ \ P = p_i+p_f 
\end{equation}
and $p_{i(f)}$ is the four-momentum of the initial (final) baryon.

To this end, we consider the matrix element
\begin{equation}
i\bar{u}_f\sigma_{\mu\nu}u_i q^{\nu} = -\frac{1}{2} \bar{u}_f
(\gamma_{\mu}\gamma_{\nu}- \gamma_{\nu}\gamma_{\mu})u_i(p_i^{\nu}- p_f^{\nu}).  \label{eom0}  
\end{equation}
By using the eom and the relationship
\begin{equation}
(\gamma_{\mu}\gamma_{\nu}+ \gamma_{\nu}\gamma_{\mu}) = 2g_{\mu\nu}, 
\end{equation}
we get
\begin{eqnarray}
\bar{u}_f(\gamma_{\mu}\gamma_{\nu}- \gamma_{\nu}\gamma_{\mu}) u_i p_i^{\nu} = 2(m_i\bar{u}_f\gamma_{\mu}u_i-\bar{u}_fu_i p_{i\mu}),
\\
\bar{u}_f(\gamma_{\mu}\gamma_{\nu}- \gamma_{\nu}\gamma_{\mu}) u_i p_f^{\nu} = 2(\bar{u}_fu_i p_{f\mu}-m_f\bar{u}_f\gamma_{\mu}u_i).
\end{eqnarray}

Inserting these two expressions into Eq. (\ref{eom0}) yields 
\begin{equation} 
i\bar{u}_f\sigma_{\mu\nu}u_i q^{\nu} = \bar{u}_fu_i P_{\mu}- (m_i+m_f)\bar{u}_f\gamma_{\mu}u_i. \label{el}
\end{equation}
By considering the matrix element $\bar{u}_f V_{\mu} u_i$ and taking account of Eq. (\ref{el}), we get Eq. (\ref{eqv}). As far as $A_{\mu}$ is concerned, a quite analogous procedure for  leads to Eq. (\ref{eqa}). 

\newpage
\setcounter{equation}{0}
 \renewcommand\theequation{B. \arabic{equation}}
\appendix{\large \bf Appendix B}
\vskip 0.30cm
Here we verify that the formula used for the differential decay width amounts, under the suitable substitutions, to well-known expression for the muon decay, $\mu^- \to e^- \nu_{\mu} \bar{\nu_e}$. Eqs. (\ref{q2}), (\ref{ddw1}) and (\ref{modsq}) yield
\begin{equation} 
\frac{d\Gamma^{SM}_{\ell}}{dE_f} = 2m_i\frac{d\Gamma^{SM}_{\ell}}{dq^2} = \frac{|V_{cb}|^2}{2^6 \pi^3 m_i} \frac{G^2}{2} 
\int_{E_{\ell}^-}^{E_{\ell}^+} dE_{\ell} T_{SM}. \label{ddw4}
\end{equation}
Here $E_{\ell}$ and $E_f$ are respectively the energies of the final baryon and of the charged lepton in the $\Lambda_b$ rest frame, with $E_{\ell}^{\pm}$ given by Eqs. (\ref{zrs}). For the sake of simplicity, we assume the Isgur-Wise form factor; then Eq. (\ref{smc}) entails
\begin{equation} 
T_{SM} = 2^7 p_f\cdot p_{\ell} ~ p_i \cdot p ~ \zeta^2_0[\omega(E_f)], \ ~~~~ \ ~~~~ \ \omega(E_f) = \frac{E_f}{m_f}, 
\label{tsm2}
\end{equation}
with $\zeta_0[\omega(E_f)]$ given by Eqs. (\ref{iwff0}) and (\ref{iwff}). Then
\begin{equation} 
\frac{d\Gamma^{SM}_{\ell}}{dE_f} = \frac{|V_{cb}|^2}{\pi^3} G^2 \zeta^2_0[\omega(E_f)]
\int_{E_{\ell}^-}^{E_{\ell}^+} dE_{\ell} [-m_i E_{\ell}^2+ A_0(E_f) E_{\ell} + B_0(E_f)], \label{ddw5}
\end{equation}
with 
\begin{eqnarray}
A_0(E_f) &=& -2m_iE_f+\frac{1}{2}M^2+m_i^2, \ ~~~~ \ ~~~~ \ \  ~~~~ \ ~~~~ \
\\
B_0(E_f) &=& -m_iE_f^2+(m_i M_0 +\frac{1}{2}M^2)E_f +\frac{1}{2}(m_{\ell}^2m_f-M^2M_0),    
\\
M^2 &=&  m_i^2+m_f^2+m_l^2, \  ~~~~ \ ~~~~ \ \ ~~~~ \ M_0 = m_i+\frac{1}{2}m_f.
\end{eqnarray}
In order to recover the differential width of the muon decay, we substitute
\begin{equation} 
m_i \to m_{\mu}, \ ~~~~ \ ~~~~ \ m_f, ~ m_{\ell} \to 0, \ ~~~~ \ ~~~~ \ \zeta_0, ~ V_{cb} \to 1. \label{mud}
\end{equation}
Therefore Eq. (\ref{ddw5}) yields 
\begin{equation}
\frac{d\Gamma^{SM}_{\ell}}{dE_f} \to \frac{d\Gamma^{SM}_{\mu}}{dE_e} = \frac{G^2}{\pi^3} m_{\mu}
[-\frac{1}{3}\delta_3(E_e)+\frac{1}{2}A(E_e)\delta_2(E_e)-B(E_e)\delta_1(E_e)]. \label{ddw6}
\end{equation} 
Here
\begin{eqnarray}
A(E_e) &=& \frac{1}{2}(3 m_{\mu}-4 E_e), \ ~~~~ \ ~~~~ \ B(E_e) = \frac{1}{4}(2E_e^2-3 m_{\mu}E_e+ m_{\mu}^2), \label{ch1}
\\
\delta_1(E_e) &=& \frac{\sqrt{\Delta}}{q^2}, \ ~~~~ \ \delta_2(E_e) = \delta_1(E_e) \frac{b}{q^2}, \ ~~~~ \  
\delta_3(E_e) = \delta_1(E_e) \frac{b^2+q^2c}{(q^2)^2}; \label{cc2} 
\end{eqnarray}
moreover, $\Delta$, $b$ and $c$ are given by Eqs. (\ref{zrs}) to (\ref{coef3}), taking into account the 
substitutions (\ref{mud}). Substituting Eqs. (\ref{ch1}) and (\ref{cc2}) into Eq. (\ref{ddw5}), we get the energy spectrum of the electron 
emerging from the muon decay\cite{ok}:
\begin{equation} 
\frac{d\Gamma^{SM}_{\mu}}{dE_e} = \frac{m_{\mu}G^2}{12\pi^3} E_e^2(3 m_{\mu}-4 E_e). \label{ddw6}
\end{equation}

The calculation of the SM differential and decay width, Eq. (\ref{ddw5}), has been performed analytically. The partial decay width has been obtained by integrating numerically the differential one between $m_f$ and $E_f^m$, according to Eqs. (\ref{lmf}). An analogous procedure has been employed for the contributions due to new physics. To this end, the tool {\it Mathematica}\cite{mth} has been used. The same results, exposed in Tables 2 and 3, have been obtained also by means of {\it Matlab}\cite{ml}.


\vskip 1cm
 
\begin{thebibliography}{0}

                    %%%   HIGGS

\bibitem{cms} S. Chatrchyan {\it et al.}, CMS Coll.: {\it Phys. Lett. B} {\bf 716} (2012) 30
\bibitem{atl} G. Aad {\it et al.}, ATLAS Coll.: {\it Phys. Lett. B} {\bf 716} (2012) 1

%\bibitem{pkn} M.E. Peskin: {\it Annalen der Phys.} {\bf 528} (2016) 20
%\bibitem{bwh} J.M. Butterworth: arXiv:1601.02759 [hep-ex]
%\bibitem{alt} G. Altarelli: {\it Phys. Scripta} {\bf T158} (2013) 014011

                %%% EXPERIMENTS                 
 
\bibitem{bb1} J.P. Lees {\it et al.}, BaBar Coll.: {\it Phys. Rev. Lett.} {\bf 109} (2012) 101802
\bibitem{bb2} J.P. Lees {\it et al.}, BaBar Coll.: {\it Phys. Rev. D} {\bf 88} (2013) 072012   
\bibitem{bel1} M. Huschle {\it et al.}, Belle Coll.: {\it Phys. Rev. D} {\bf 92} (2015) 072014
\bibitem{bel2} Y. Sato {\it et al.}, Belle Coll.: {\it Phys. Rev. D} {\bf 94} (2016) 072007
\bibitem{bel3} S. Hirose {\it et al.}, Belle Coll.: {\it Phys. Rev. Lett.} {\bf 118} (2017) 211801
\bibitem{lhcb3} R. Aaij {\it et al.}, LHCb Coll.: {\it Phys. Rev. Lett.} {\bf 115} (2015) 111803
\bibitem{lhcb1} R. Aaij {\it et al.}, LHCb Coll.: {\it Phys. Rev. Lett.} {\bf 111} (2013) 191801
\bibitem{lhcb2} R. Aaij {\it et al.}, LHCb Coll.: {\it Phys. Rev. Lett.} {\bf 113} (2014) 151601

                %%% SM CALCULATIONS 
  
\bibitem{kam} J.F. Kamenik and F. Mescia: {\it Phys. Rev. D} {\bf 78} (2008) 014003 
\bibitem{faj} S. Fajfer {\it et al.}: {\it Phys. Rev. D} {\bf 85} (2012) 094025
\bibitem{bai} J.A. Bailey {\it et al.}, Fermilab Lattice and Milc Coll.: {\it Phys. Rev. Lett.} {\bf 109} (2012) 071802; 
{\it Phys. Rev. D} {\bf 92} (2015) 034506

               %%% MODEL INDEP. ANALYSES

\bibitem{lee2} J.P. Lee: {\it Phys. Lett. B} {\bf 526} (2002) 61
\bibitem{faj2} S. Fajfer {\it et al.}: {\it Phys. Rev. Lett.} {\bf 109} (2012) 161801
\bibitem{dt1} A. Datta {\it et al.}: {\it Phys. Rev. D} {\bf 86} (2012) 034027
\bibitem{tw} M. Tanaka and R. Watanabe: {\it Phys. Rev. D} {\bf 87} (2013) 034028
\bibitem{bcd} P. Biancofiore {\it et al.}: {\it Phys. Rev. D} {\bf 87} (2013) 074010
\bibitem{iv}  M.A. Ivanov {\it et al.}: {\it Phys. Rev. D} {\bf 94} (2016) 094028;
{\it Phys. Rev. D} {\bf 95} (2017) 036021
\bibitem{ch17} D. Choudhury {\it et al.}: {\it Phys.Rev. D} {\bf 95} (2017) 035021 
\bibitem{bht}  S. Bhattacharya {\it et al.}: {\it Phys. Rev. D} {\bf 93} (2016) 034011; {\it Phys. Rev. D} {\bf 95} (2017) 075012
\bibitem{dln} L. Di Luzio and M. Nardecchia: {\it Eur. Phys. Jou. C} {\bf 77} (2017) 536
\bibitem{bln} F.U. Bernlochner {\it et al.}: {\it Phys. Rev. D} {\bf 95} (2017) 115008

               %%% 2HD MODELS

\bibitem{dhg} L. Dhargyal: {\it Phys. Rev. D} {\bf 93} (2016) 115009
\bibitem{lee} J.P. Lee: {\it Phys. Rev. D} {\bf 96} (2017) 055005
\bibitem{igt} S. Iguro and K. Tobe: {\it Nucl. Phys. B} {\bf 925} (2017) 560
\bibitem{chn} C.-H. Chen and T. Nomura: {\it Eur. Phys. Jou. C} {\bf 77} (2017) 631

               %%% LQ MODELS

\bibitem{sk} Y. Sakaki {\it et al.}: {\it Phys. Rev. D} {\bf 88} (2013) 094012
\bibitem{frt} M. Freytsis {\it et al.}: {\it Phys. Rev. D} {\bf 92} (2015) 054018
\bibitem{bb} R. Barbieri {\it et al.}: {\it Eur. Phys. Jou. C} {\bf 77} (2017) 8
\bibitem{crv} A. Crivellin {\it et al.}: {\it JHEP} {\bf 1709} (2017) 040
\bibitem{cai} Y. Cai {\it et al.}: {\it JHEP} {\bf 1710} (2017) 047
\bibitem{btz} D. Buttazzo {\it et al.}: {\it JHEP} {\bf 1711} (2017) 044
\bibitem{bdn} M. Bordone {\it et al.}: {\it Phys. Rev. D} {\bf 96} (2017) 015038

                %%% L-R MODELS

\bibitem{hv1} X.-G. He and G. Valencia: {\it Phys. Rev. D} {\bf 87} (2013) 014014; {\it Phys. Lett. B} {\bf 779} (2018) 52
\bibitem{alt} W. Altmannshofer {\it et al.}: {\it Phys. Rev. D} {\bf 96} (2017) 095010


                 %%% OTHER MODELS

\bibitem{bsw} A. Biswas {\it et al.}: {\it Phys.Rev. D} {\bf 97} (2018) 035019  %%% EXTRA DIMENSIONS



                 %%% CONNECTION TO g-2

\bibitem{cg} C.-H. Chen and C.-Q. Geng: {\it Phys. Rev. D} {\bf 71} (2005) 077501

                 


                 %%% Lambda_b --> Lambda ll 

\bibitem{dm} W. Detmold and S. Meinel: {\it Phys. Rev. D} {\bf 93} (2016) 074501 
\bibitem{gu} T. Gutsche {\it et al.}: {\it Phys. Rev. D} {\bf 87} (2013) 074031

                 %%% Lambda_b --> Lambda_c tau nu

\bibitem{swd} S. Shivashankara {\it et al.}: {\it Phys. Rev. D} {\bf 91} (2015) 115003 
\bibitem{gu1} T. Gutsche {\it et al.}: {\it Phys. Rev. D} {\bf 91} (2015) 119907 
\bibitem{dut} R. Dutta: {\it Phys. Rev. D} {\bf 93} (2016) 054003
\bibitem{hgi} N. Habyl {\it et al.}: {\it Int. Jou. Mod. Phys. Conf. Ser.} {\bf 39} (2015) 1560112
\bibitem{dsa} E. Di Salvo and Z.J. Ajaltouni: {\it Mod. Phys. Lett.
A} {\bf 32} (2017) 1750043 
\bibitem{lyz} X.Q. Li {\it et al.}: {\it JHEP} {\bf 1702} (2017) 068
\bibitem{dt2} A. Datta {\it et al.}: {\it JHEP} {\bf 1708} (2017) 131

            %%%%%%% B_c --> J\psi l nu

\bibitem{dub} R. Dutta and A. Bohl: {\it Phys. Rev. D} {\bf 96} (2017) 076001

            %%%% AVERAGE VALUES OF DOUBLE RATIOS

            %%%%% FORM FACTORS

\bibitem{dm1} W. Detmold {\it et al.}: {\it Phys. Rev. D} {\bf 92} (2015) 034503
\bibitem{pv} M. Pervin {\it et al.}: {\it Phys. Rev. C} {\bf 72} (2005) 035021
\bibitem{guf} T. Gutsche {\it et al.}: {\it Phys. Rev. D} {\bf 90} (2014) 114033
\bibitem{fgk} R.N. Faustov and V.O. Galkin: {\it Phys. Rev. D} {\bf 94} (2016) 073008
\bibitem{klw} H.-W. Ke {\it et al.}: {\it Phys. Rev. D} {\bf 77} (2008) 014020
\bibitem{lhcb4} R. Aaij {\it et al.}, LHCb Coll.: {\it Phys. Rev. D} {\bf 96} (2017) 112005
\bibitem{dec} R.S.M. De Carvalho {\it et al.}: Phys. Rev. D {\bf 60} (1999) 034009 
\bibitem{azs} K. Azizi and J.Y. Sungu: {\it Phys. Rev. D} {\bf 97} (2018) 074007


              %%% SIMULT. EXPL. OF ANOMALIES - 1


\bibitem{choh} D. Choudhury {\it et al.}: {\it Phys. Rev. Lett.} {\bf 119} (2017) 151801 
\bibitem{hf16} Y. Amhis {\it et al.}, HFLAV coll.: {\it Eur. Phys. Jou. C} {\bf 77} (2017) 895


           %%% PARTICLE DATA GROUP
   
\bibitem{pdg} C. Patrignani {\it et al.}: {\it Chin. Phys. C} {\bf 40} (2016) 100001


             %%%% ASYMMETRIES 

\bibitem{ddt} M. Duraisamy and A. Datta: {\it JHEP} {\bf 1309} (2013) 059 


             %%%% GLASHOW: B --> K \ell \ell 

\bibitem{gls} S. Glashow {\it et al.}: {\it Phys. Rev. Lett.} {\bf 114} (2015) 091801

            %%% SIMULT. EXPL. OF ANOMALIES - 2

\bibitem{bht2} B. Bhattacharya {\it et al.}: {\it Phys. Lett. B} {\bf 742} (2015) 370


             %%%%% ALTERNATIVE ANALYSIS

\bibitem{clj} A. Celis {\it et al.}: {\it Phys. Lett. B} {\bf 771} (2017) 168

             %%%%% POLARIZATION

\bibitem{kum} A.K. Alok {\it et al.}: {\it Phys. Rev. D} {\bf 95} (2017) 115038 

              %%%% MUON DECAY

\bibitem{ok} L.B. Okun: {\it Leptons and Quarks}, Elsevier Science Publishers B.V., 1982, 1984

              %%%% MATHEMATICAL TOOLS

\bibitem{mth} Wolfram Research, Inc., Mathematica, Version 11.3, Champaign, IL (2018)
\bibitem{ml}  MATLAB Release 2016b, The MathWorks, Inc., Natick, Massachusetts, United States
 


%%%%%%%%%%%%%%%%%%%%%%%%%%%%%%%%%%%%%%%%%%%%%%%%%

%\bibitem{dlm} W. Detmold {\it et al.}: {\it Phys. Rev. D} {\bf 87} (2013) %074502 
%\bibitem{jjo} F. Jugeau {\it et al.}: {\it Phys. Rev. D} {\bf 86} (2012) %014002
%\bibitem{va} G. Valencia: {\it Phys. Rev. D} {\bf 39} (1989) 3339
%\bibitem{gv} E. Golowich and G. Valencia: {\it Phys. Rev. D} 
%{\bf 40} (1989) 112
%\bibitem{wf} L. Wolfenstein: {\it Phys. Rev. D} {\bf 43} (1991) 151
%\bibitem{bl} W. Bensalem and D. London: {\it Phys. Rev. D} {\bf 64} %(2001) 116003
%\bibitem{bdl} W. Bensalem, A. Datta and D. London: {\it Phys. Rev. D} %{\bf 66} (2002) 094004
%\bibitem{arg} S. Arunagiri and C.Q. Geng: {\it Phys. Rev. D} {\bf 69} %(2004) 017901
%\bibitem{gr} M. Gronau and J.L. Rosner: {\it Phys. Rev. D} {\bf 84} %(2011) 096013
%\bibitem{gr1} M. Gronau and J.L. Rosner: {\it Phys. Lett. B} 
%{\bf 749} (2015) 104
%\bibitem{ajd} Z.J. Ajaltouni and E. Di Salvo: {\it Int. Jou. Mod. Phys. %E} {\bf 22} (2013) 1330006      
%\bibitem{dc14} See refs. 14 of \cite{ddg}
%\bibitem{as} W. Altmannshofer and D.M. Straub: {\it Eur. Phys. Jou. C} %{\bf 75} (2015) 382
%\bibitem{iw} N. Isgur and M. Wise: {\it Nucl. Phys. B} {\bf 348} (1991) %276
%\bibitem{yor} A. Le Yaouanc {\it et al.}: {\it Phys. Rev. D} {\bf 79} %(2009) 014023
%bibitem{gkl} P. Guo {\it et al.}: {\it Phys. Rev. D} {\bf 75} (2007) %054017
%\bibitem{kkk} B. Koenig {\it et al.}: {\it Phys. Rev. D} {\bf 56} (1997) %4282 
%\bibitem{dlm1} W. Detmold {\it et al.}: {\it Phys. Rev. D} {\bf 92} %(2015) 034503
%\bibitem{klld} X.-W. Kang {\it et al.}: {\it Int. Jou. Mod. Phys. A} %{\bf 26} (2011) 2523


\end{thebibliography}
\end{document}